\documentclass[aps,prd,preprint,superscriptaddress,nofootinbib]{revtex4-1}
\usepackage{braket}
\usepackage{xargs}
\usepackage{mathtools}
\usepackage[english]{babel}
\usepackage[utf8]{inputenc}
\usepackage{amsmath,amssymb,braket,slashed,mathrsfs}
\usepackage{pifont}
\usepackage{graphicx}
\usepackage{tikz}
\usetikzlibrary{shapes,arrows,positioning}
\tikzset{decision/.style={diamond, draw, fill=blue!20, text width=4.5em, text badly centered, inner sep=0pt}}
\tikzset{block/.style={rectangle, draw, fill=blue!20, text width=10em, text centered, rounded corners, minimum width=3.5cm}}
\tikzset{block1/.style={rectangle, draw, fill=blue!20, text width=18.5em, text centered, rounded corners, minimum width=3.5cm}}
\tikzset{line/.style={draw, -latex, thick}}
\usepackage{color,hyperref}
\usepackage{multirow,array}

\usepackage{cleveref}
\newcommand{\ba}{\begin{eqnarray}}
\newcommand{\ea}{\end{eqnarray}}
\newcommand{\be}{\begin{equation}}
\newcommand{\ee}{\end{equation}}

\newcommand{\nn}{\nonumber}

\newcommand{\lat}{\mathrm{lat}}

\newcommand{\QEDL}{{\text{QED}_{\mathrm{L}}}}
\newcommand{\IVR}{\text{IVR}}

\newcommand{\innovation}{Collaborative Innovation Center of Quantum Matter, Beijing 100871, China}
\newcommand{\chep}{Center for High Energy Physics, Peking University, Beijing 100871, China}
\newcommand{\pkuphy}{School of Physics, Peking University, Beijing 100871,
China}
\newcommand{\KeyLab}{State Key Laboratory of Nuclear Physics and Technology,
Peking University, Beijing 100871, China}
\newcommand{\Uconn}{Department of Physics, University of Connecticut, Storrs, CT 06269, USA}
\newcommand{\RBRC}{RIKEN-BNL Research Center, Brookhaven National Laboratory, Building 510, Upton, NY 11973}

\begin{document}
\title{Long-distance Contributions to Neutrinoless Double Beta Decay $\pi^- \to\pi^+ e e $}

\author{Xin-Yu Tuo}\affiliation{\pkuphy}
\author{Xu~Feng}\email{xu.feng@pku.edu.cn}\affiliation{\pkuphy}\affiliation{\innovation}\affiliation{\chep}\affiliation{\KeyLab}
\author{Lu-Chang Jin}\email{ljin.luchang@gmail.com}\affiliation{\Uconn}\affiliation{\RBRC}
\pacs{PACS}

\date{\today}

\begin{abstract}

Neutrinoless double beta decay, if detected, would prove that neutrinos are
    Majorana fermions and provide the direct evidence for lepton number
    violation.
    If such decay would exist in nature, then $\pi^-\pi^-\to ee$ and
    $\pi^-\to\pi^+ ee$ (or equivalently $\pi^-e^+\to\pi^+ e^-$) are the two
    simplest processes accessible via first-principle lattice QCD calculations.
    In this work, we calculate the long-distance
    contributions to the $\pi^-\to\pi^+ee$ transition amplitude
    using four ensembles at the physical pion mass with various volumes and
    lattice spacings. We adopt the infinite-volume reconstruction method~\cite{Feng:2018qpx} to
    control the finite-volume effects arising from the (almost) massless neutrino. 
    Providing the lattice QCD inputs for chiral perturbation
    theory, we obtain the low energy constant
    $g_\nu^{\pi\pi}(m_\rho)=-10.89(28)_\text{stat}(74)_\text{sys}$, which is close
    to $g_\nu^{\pi\pi}(m_\rho)=-11.96(31)_\text{stat}$ determined from the
    crossed-channel $\pi^-\pi^-\to ee$ decay~\cite{Feng:2018pdq}.

\end{abstract}

\maketitle
   
\section{Introduction}

Observation of neutrinoless double beta ($0\nu2\beta$) decays would prove neutrinos 
as Majorana fermions and lepton number violation in nature. 
The light-neutrino exchange is the most widely discussed mechanism 
to explain $0\nu2\beta$ decays. Under this mechanism, 
the decay amplitude is proportional to the effective neutrino mass $m_{\beta\beta}$ 
and thus the detection of $0\nu2\beta$ decay would provide the information 
about the absolute neutrino mass,
while the neutrino oscillation experiments are only sensitive to the mass differences among neutrinos.

Due to its importance, the detection of $0\nu2\beta$ decays is being pursued by many experiments around the 
world~\cite{Gando:2012zm,Agostini:2013mzu,Albert:2014awa,Andringa:2015tza,KamLAND-Zen:2016pfg,Elliott:2016ble,Agostini:2017iyd,Azzolini:2018dyb,Aalseth:2017btx,Albert:2017owj,Alduino:2017ehq,Agostini:2018tnm}.
Current experimental measurements of the decay's half-lives $T_{1/2}^{0\nu}$
have reached
the level of $T_{1/2}^{0\nu}>1.07\times10^{26}$ yr for $^{126}$Xe~\cite{KamLAND-Zen:2016pfg},
with a new generation of ton-scale experiments aiming for the level of sensitivity
improved by 1 or 2 orders of magnitude.

On the theoretical side, current knowledge of second-order weak-interaction nuclear matrix elements needs to be improved, 
as various nuclear models lead to discrepancies on the order of 100\%~\cite{Engel:2016xgb}.
The interpretation of $0\nu2\beta$ experiments relies on
reliable calculations of the nuclear matrix elements, with robust uncertainty estimation.
While the heavy nuclei system is beyond the capability of the current lattice
QCD calculation, computations of the double beta decay for a light nuclei system
shall be feasible~\cite{Shanahan:2017bgi,Tiburzi:2017iux}.
The lattice results are required as inputs to
determine the relevant low energy constants in the effective field
theory~\cite{Cirigliano:2017tvr,Cirigliano:2018yza,Cirigliano:2018hja,Cirigliano:2019vdj},
with which the nuclear matrix elements for heavy nuclei system can be calculated.

Without the signal-to-noise-ratio problem, the decay channels $\pi^-\pi^-\to ee$
and $\pi^-\to\pi^+ee$ serve as an ideal laboratory to perform a lattice
QCD study of the $0\nu2\beta$
decay and to test the prediction from effective field theory.
Our exploratory study~\cite{Feng:2018pdq} has demonstrated the possibility of a first-principles 
calculation of the $\pi^-\pi^-\to ee$ decay, where we obtained the decay
amplitude
\be
\frac{\mathcal{A}(\pi\pi\to ee)}{\mathcal{A}^{LO}}\bigg|_{m_\pi=420\,\,
\mathrm{MeV}}=0.759(6)_{\text{stat}},
\quad \frac{\mathcal{A}(\pi\pi\to ee)}{\mathcal{A}^{LO}}\bigg|_{m_\pi=140\,\,
\mathrm{MeV}}=0.910(3)_{\text{stat}},
\ee
with $\mathcal{A}^{LO}$ the leading-order prediction from chiral perturbation
theory ($\chi$PT).
By putting the amplitude into the $\chi$PT formula~\cite{Cirigliano:2017tvr}
\be
\label{eq:ChPT1}
\frac{\mathcal{A}(\pi\pi\to ee)}{\mathcal{A}^{LO}}=1-\frac{m_\pi^2}{(4\pi
F_\pi)^2}\left(3\log\frac{\mu^2}{m_\pi^2}+\frac{7}{2}+\frac{\pi^2}{4}+\frac{5}{6}g_\nu^{\pi\pi}(\mu)\right),
\ee
we obtain the low energy constant $g_\nu^{\pi\pi}(\mu)$ at $\mu=m_\rho=775$ MeV
\be
\label{eq:LEC1}
g_\nu^{\pi\pi}(m_\rho)\Big|_{m_\pi=420\,\,\mathrm{MeV}}=-8.50(9)_\text{stat},\quad
g_\nu^{\pi\pi}(m_\rho)\Big|_{m_\pi=140\,\,\mathrm{MeV}}=-11.96(31)_\text{stat},
\ee
where the uncertainties are statistical only. The two values of $g_\nu^{\pi\pi}$
differ by $\sim30\%$.
This can be accounted by the systematic effects in the lattice
calculation such as finite-volume effects and lattice artifacts, as well as
higher-order truncation effects from $\chi$PT.

The lattice QCD calculations of $\pi^-\to\pi^+ee$ decay have been first carried
out by CalLat Collaboration~\cite{Nicholson:2018mwc}
for the short-distance contribution and NPLQCD Collaboration~\cite{Detmold:2018zan} for the
long-distance contribution. 
While the vanishing phase space does not allow the $\pi^-\to\pi^+ee$ decay
happen in nature (This problem does not exist for the $K^-\to\pi^+ee$ decay,
which is proposed by Ref.~\cite{Liao:2019tep}), the hadronic matrix element is well defined within the
Standard Model and is equivalent to the one from $\pi^- e^+\to\pi^+e^-$
scattering, where $\pi^\pm$ and $e^\pm$ carry zero spatial momentum.
As the crossed-channel analog to the $\pi^-\pi^-\to ee$ decay, the process of $\pi^-\to\pi^+ee$ can
be combined together with $\pi^-\pi^-\to ee$ and serves as a cross-check 
for the prediction from $\chi$PT. Since in $\pi^-\to\pi^+ee$ decay the initial
and final state only involves a single stable hadron, the study of the
finite-volume effects is simplified. For example, we can adopt a newly developed
technique called infinite-volume reconstruction~\cite{Feng:2018qpx} to determine the decay
amplitude, where the
finite-volume effects are exponentially suppressed 
even a massless neutrino propagator is included in the lattice calculation.
Using four ensembles with different volumes and lattice spacings, we obtain the decay amplitude as 
\be
\frac{\mathcal{A}(\pi^-\to\pi^+ee)}{\mathcal{A}^{LO}}\bigg|_{m_\pi=140\,\,\mathrm{MeV}}=
1.1045(34)_\text{stat}(74)_\text{sys},
\ee
where the first uncertainty is statistical and the second one is an estimation 
for both finite-volume effects and lattice artifacts. Using the $\chi$PT formula
for $\pi^-\to\pi^+ee$ decay~\cite{Cirigliano:2017tvr}
\begin{equation}
\label{eq:ChPT2}
\begin{split}
    \frac{\mathcal{A}(\pi^-\to\pi^+ee)}{\mathcal{A}^{LO}}=1+\frac{m_\pi^2}{(4\pi
    F_\pi)^2}\left(3\log\frac{\mu^2}{m_\pi^2}+6+\frac{5}{6}
    g_\nu^{\pi\pi}(\mu)\right)
\end{split}
\end{equation}
we obtain the low energy constant
\be
\label{eq:LEC2}
g_{\nu}^{\pi\pi}(m_\rho)\Big|_{m_\pi=140\,\,\mathrm{MeV}}=-10.89(28)_\text{stat}(74)_\text{sys}.
\ee
Although the functional forms of $\chi$PT formulae~(\ref{eq:ChPT1}) and
(\ref{eq:ChPT2}) are
quite different, the results for $g_{\nu}^{\pi\pi}$ given in (\ref{eq:LEC1}) and
(\ref{eq:LEC2}) are close to each other, demonstrating the success of $\chi$PT
prediction.

\section{Calculation of $0\nu 2\beta$ process: $\pi^- \to\pi^+ e e $}

The decay amplitude of a general $0\nu 2\beta$ process $I(p_I)\to F(p_F)e(p_1)e(p_2)$ 
can be written as
\begin{equation}
\mathcal{A}=\langle F,e_1,e_2|\mathcal{H}_\text{eff}|I\rangle.
\end{equation}
Here we use $e_{1,2}$ to specify the electron state carrying momentum $p_{1,2}$.
The second-order weak effective Hamiltonian is defined as
\begin{equation}
\begin{split}
\mathcal{H}_\text{eff}&=\frac{1}{2!}\int d^4 x\,
    T[\mathcal{L}_\text{eff}(x)\mathcal{L}_\text{eff}(0)],\\
\mathcal{L}_\text{eff}&=2\sqrt{2}G_FV_{ud}(\bar{u}_L\gamma_\mu
    d_L)(\bar{e}_L\gamma_\mu\nu_{eL}).\\
\end{split}
\end{equation}
Here $G_F$ is the Fermi constant and $V_{ud}$ is the CKM matrix element. The
left-handed fermion fields are defined as 
$\psi_L=P_L \psi$, $\bar{\psi}_L=\bar{\psi}P_R$ (for
$\psi=u,d,e,\nu_e$)
with projectors $P_{L,R}=(1\mp\gamma_5)/2$. 

The effective Hamiltonian can be written as a product of hadronic and leptonic factors
\be
\mathcal{H}_\text{eff}=H_{\mu\nu}(x)L_{\mu\nu}(x),
\ee
where the hadronic factor $H_{\mu\nu}(x)$ is defined as $H_{\mu\nu}(x)=T[J_{\mu L}(x)J_{\nu L}(0)]$
with $J_{\mu L}(x)=\bar{u}_L\gamma_\mu d_L(x)$.
Under the mechanism that $0\nu2\beta$ decays are mediated by the exchange of
light Majorana neutrinos, the leptonic factor
$L_{\mu\nu}(x)$ is given by
\be
L_{\mu\nu}(x)=-4G_F^2V^2_{ud}m_{\beta\beta}\,S_0(x)\,\bar{e}_L(x)\gamma_\mu\gamma_\nu
e_L^c(0),
\ee
where $S_0(x)=\int\frac{d^4q}{(2\pi)^4}\frac{e^{iqx}}{q^2}$ denotes a massless scalar
propagator and
the effective neutrino mass $m_{\beta\beta}=|\sum_i U_{ei}^2m_i|$ combines the
neutrino masses $m_i$ and the elements $U_{ei}$ of the Pontecorvo-Maki-Nakagawa-Sato (PMNS) matrix.
The charge conjugate of a fermionic field $\psi$ is given as
$\psi^c=C\bar{\psi}^T=\gamma_4\gamma_2\bar{\psi}^T$.

\subsection{Decay amplitude of $\pi^- \to\pi^+ e e $}

For specific process $\pi^- \to\pi^+ e e $, with two electrons carrying
vanishing momenta, the decay amplitude in Minkowski space-time becomes
\begin{equation}
    \label{eq:integral}
    \mathcal{A}^{M}=-2T_\text{lept}\int d^4x\,H^{M}(x)S_0^{M}(x)
\end{equation}
where leptonic part is factorized in
$T_\text{lept}=4G_F^2V_{ud}^2m_{\beta\beta}\bar{e}_L(p_1)e^c_L(p_2)$. The
superscript {\em M} denotes the Minkowski space-time. The factor of
$2$ comes from interchange of electrons. The hadronic function 
is defined by
\begin{equation}
\label{eq:had_func}
H^M(x)=\langle \pi^+|T\{J_{\mu L}^M(x)J_{\mu L}^M(0)\}|\pi^-\rangle.
\end{equation}

This calculation is similar to the calculation of QED corrections to self energy using
Feynman gauge.
We can adopt the infinite-volume reconstruction (IVR) method proposed in
Ref.~\cite{Feng:2018qpx} to compute the $\pi^-\to\pi^+ee$ transition amplitude.

It should be noted that the hadronic function receives contribution from the
vacuum state, which is lighter than the single pion state.
In the Euclidean space-time, the hadronic function would grow exponentially
as the time separation between the two current operators increases.
To reproduce the amplitude in Eq.~(\ref{eq:integral}), one needs to treat the
vacuum state properly as we will describe later.

In the following sections, we will first introduce our approach to calculate
the Euclidean space-time hadronic function $H(x)$ on lattice (For simplicity, we
have left out the superscript of {\em E} for Euclidean space-time)
and connect it with the Minkowski space-time integral, Eq. (\ref{eq:integral}).
Then we use two different methods, QED$_\mathrm{L}$ and IVR, to calculate the
integral in Eq.~(\ref{eq:integral}).
The results from the two methods are compared and discussed later.   

\subsection{Calculation of the hadronic function}
In order to calculate the Euclidean space-time hadronic function on lattice,
we define the following four point correlation function
 \begin{equation}
 \begin{split}
 C(t_f,x,y,t_i)=\langle\phi_{\pi^+}(t_f)J_{\mu L}(x)J_{\mu L}(y)\phi^\dagger_{\pi^-}(t_i)\rangle
 \end{split}
 \end{equation} 
with wall-source pion interpolating operators $\phi^\dagger_{\pi^-}$ and
$\phi_{\pi^+}$. Here the time slices
$t_i$ and $t_f$ are chosen as
\begin{equation}
\begin{split}
    t_i=\operatorname{min}(x_t,y_t)-\Delta T,\quad t_f=\operatorname{max}(x_t,y_t)+\Delta T
\end{split}
\end{equation}
with sufficiently large $\Delta T$ for the ground-state saturation. Since the
wall-source operators have a good overlap with the $\pi$ ground state, we find
the ground-state saturation for $\Delta T\gtrsim 1$ fm at the physical pion
mass.
For
the ensembles used in this work, the values of $\Delta T$ are chosen
conservatively and listed in
Table~\ref{tab:ensemble_parameter}.

\begin{table}[htbp]
	\small
	\centering
	\begin{tabular}{ccccccc}
		\hline
        Ensemble  & $m_\pi$ [MeV] & $L^3\times T$ & $a^{-1}$ [GeV]& $N_{conf}$ &
        $m_\pi L$ & $\Delta T/a$ \\
		\hline
        24D  & 142 & $24^3\times 64$ & $1.015$ & 91 & 3.3 & 8  \\
		\hline
        32D  & 142 & $32^3\times 64$ & $1.015$ & 56 & 4.5 & 8 \\
		\hline
        32D-fine & 143 & $32^3\times64$ & $1.378$ & 24 & 3.3 & 10 \\
        \hline
        48I & 139 & $48^3\times 96$ & $1.73$ & 34 & 3.9 & 12 \\
    \hline
    \end{tabular}%
    \caption{Ensembles used in this work are generated by the RBC/UKQCD collaborations~\cite{Blum:2014tka,lattice2018:robert}. We list the pion mass $m_\pi$, the space-time volume $L^3\times T$, the lattice
    spacing $a$, the number, $N_{\mathrm{conf}}$,
    of configurations used, the values of $m_\pi L$ and the
    time separation, $\Delta T$, used for the $\pi$
    ground-state saturation.}
    \label{tab:ensemble_parameter}%
\end{table}%

\begin{figure}[htbp]
	\centering
	\includegraphics[width=0.8\textwidth]{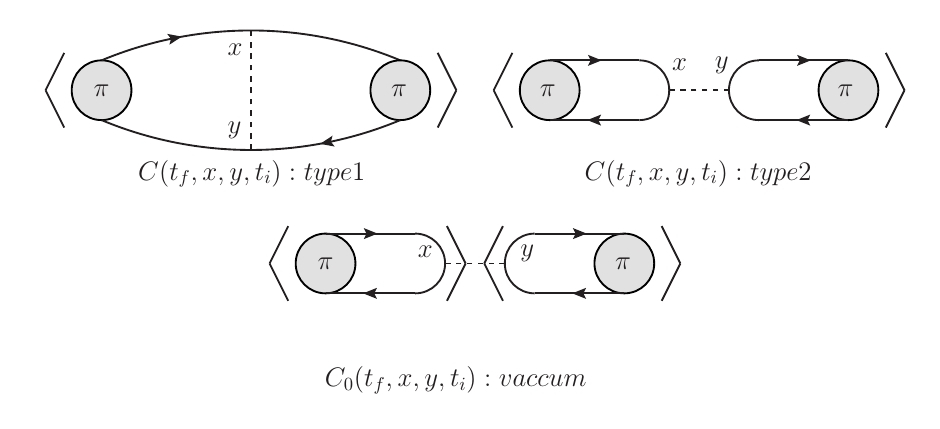}
    \caption{Contractions of correlation function $C(t_f,x,y,t_i)$ and
    $C_0(t_f,x,y,t_i)$.}
    \label{fig:contraction}
\end{figure}

The Euclidean space-time hadronic function is given by:
\begin{equation}
 \begin{split}
 H(x-y)=V\frac{C(t_f,x,y,t_i)}{C_\pi(t_f,t_i)},~~~~C_\pi(t_f,t_i)=\langle\phi_{\pi^-}(t_f)\phi^\dagger_{\pi^-}(t_i)\rangle
 \end{split}
\end{equation}
where $V=L^3$ is the spatial-volume factor.
In $H(x-y)$ the vacuum-intermediate-state contribution from
      $\langle\pi^+|J_{\mu L}(x)|0\rangle\langle0|J_{\mu L}(y)|\pi^-\rangle$
      leads to an exponentially growing factor $e^{m_\pi (t_x-t_y)}$ in Euclidean
      correlator when $t_x-t_y$ increases.
Here, we define the subtracted Euclidean space-time hadronic function:
\begin{equation}
\begin{split}
    H'(x)=H(x)-H_0(x),
\end{split}
\end{equation}  
with $H_0(x)$ defined as 
\begin{equation}
\begin{split}
    &H_0(x-y)=V\frac{C_0(t_f,x,y,t_i)}{C_\pi(t_f,t_i)},\\
    &C_0(t_f,x,y,t_i)=\langle\phi_{\pi^+}(t_f)J_{\mu L}(x)\rangle\langle J_{\mu
    L}(y)\phi^\dagger_{\pi^-}(t_i)\rangle+\langle\phi_{\pi^+}(t_f)J_{\mu
    L}(y)\rangle\langle J_{\mu L}(x)\phi^\dagger_{\pi^-}(t_i)\rangle.
\end{split}
\end{equation}
The contractions of correlation function $C(t_f,x,y,t_i)$ are given by the
{\em type1} and {\em type2} diagrams in Fig.~\ref{fig:contraction}, while the
contractions of
$C_0(t_f,x,y,t_i)$ are given by the {\em vacuum} diagram,
where the two hadronic parts are only connected through a neutrino propagator.
It should be noted that either $x_t < y_t$ or $x_t > y_t$ are possible. After
the subtraction, we remove the unphysical, exponentially-growing contributions.
 
In Minkowski space-time,
the amplitude contributed by the {\em vacuum} diagram, $\mathcal{A}_0^M$, 
are known analytically and takes the simple form as
\begin{equation}
    \label{eq:I_0}
    \mathcal{A}_0^M=-2T_\text{lept} F_\pi^2,
\end{equation}
where the decay constant $F_\pi$ is defined as
\begin{equation}
    \langle 0 |(\bar{u}\gamma_\mu \gamma_5 d)^M|\pi^-(p)\rangle=
    i\sqrt{2} p_\mu F_\pi.
\end{equation}
Comparing $\mathcal{A}_0^M$ with the $\chi$PT formula~\cite{Cirigliano:2017tvr}
\begin{equation}
\begin{split}
    \mathcal{A}_{\chi PT}=2 T_\text{lept}F_\pi^2\left[1+\frac{m_\pi^2}{(4\pi
    F_\pi)^2}\left(3\log\frac{\mu^2}{m_\pi^2}+6+\frac{5}{6}
    g_\nu^{\pi\pi}(\mu)\right)\right]
\end{split}
\end{equation}
we can find that $\mathcal{A}_0^M$ is just the leading order term in
$\chi$PT. 

In the Euclidean space-time, although $H_0$ function cannot reproduce the
physical vacuum contribution $\mathcal{A}_0^M$, by removing it and using the hadronic function $H'(x)$ as inputs, 
we can obtain the subtracted amplitude,
$\mathcal{A}'=\mathcal{A}^M-\mathcal{A}_0^M$, which includes the higher-order
$\chi$PT contributions.
Through out the paper, we will
calculate the dimensionless, normalized amplitude $A$
\begin{equation}
\label{eq:norm_amplitude}
\begin{split}
    A=\frac{\mathcal{A}'}{\mathcal{A}_0^M}\equiv -\frac{1}{F_\pi^2}\int d^4 x\,
    H'(x)S_0(x),
\end{split}
\end{equation}
which can be used to determine low energy constant
$g_\nu^{\pi\pi}$ via
\be
\label{eq:A_ChPT}
A=\frac{m_\pi^2}{(4\pi
    F_\pi)^2}\left(3\log\frac{\mu^2}{m_\pi^2}+6+\frac 56
    g_\nu^{\pi\pi}(\mu)\right)+O\left(\frac{m_\pi^2}{(4\pi F_\pi)^2}\right)^2.
\ee

\subsection{Lattice setup}

We use four ensembles at the physical pion mass generated by the RBC and UKQCD
Collaborations~\cite{Blum:2014tka,lattice2018:robert}. 
The corresponding parameters are listed in
Table~\ref{tab:ensemble_parameter}. 
For the ensembles 24D and 32D, lattice spacings are the same but the lattice
volumes are different. It allows us to study the finite-volume effects. The
ensemble 48I,
32D-fine and 24D have different lattice spacings but similar volumes.
These ensembles
provide us the information to examine the lattice artifacts. Note that the 48I
uses Iwasaki gauge action in the simulation while the other three ensembles use
Iwasaki+DSDR action.

We produce wall-source light-quark propagators on all time slices and make use of the time translation
invariance to average the correlator over all $T$ time translations.
We have used AMA~\cite{Shintani:2014vja}
and low modes deflation with compressed eigen-vectors~\cite{Clark:2017wom}.
These techniques have greatly reduced the computational cost for generating propagators.
The correlators for the {\em type1} and {\em type2} diagrams in Fig.~\ref{fig:contraction}
are given by
\ba
C_{type1}(x-y)&=&\operatorname{Tr}\left[\gamma_5S(t_f;x)\gamma_{\mu
L}S(x;t_i)\,\gamma_5S(t_i;y)\gamma_{\mu
L}S(y;t_f)\right]+\{x\leftrightarrow y\}
\nn\\
C_{type2}(x-y)&=&-\operatorname{Tr}\left[\gamma_5S(t_i;x)\gamma_{\mu
L}S(x;t_i)\right]\operatorname{Tr}\left[\gamma_5S(t_f;y)\gamma_{\mu
L}S(y;t_f)\right]+\{x\leftrightarrow y\},
\ea
where $S$ the light-quark propagator.
We can write the above correlator in a general form of
\be
C(x-y)=H_1(x)H_2(y).
\ee
Such form allows us to obtain a spatial volume average of $C(x)$
by using the double Fourier transformation. We have
\be
\begin{split}
    C(x)&=\frac{1}{V}\sum_{\vec{y}}H_1(x+y)H_2(y)\\
    &=\frac{1}{V}\sum_{\vec{y}}\left(\frac{1}{V}\sum_{\vec{p}}\tilde{H}_1(t_x,\vec{p})e^{i\vec{p}\cdot(\vec{x}+\vec{y})}\right)
    \left(\frac{1}{V}\sum_{\vec{q}}\tilde{H}_2(t_y,\vec{q})e^{i\vec{q}\cdot\vec{y}}\right)\\
    &=\frac{1}{V}\left(\frac{1}{V}\sum_{\vec{p}}\tilde{H}_1(t_x,\vec{p})\tilde{H}_2(t_y,-\vec{p})e^{i\vec{p}\cdot\vec{x}}\right)
\end{split}
\ee
where $\tilde{H}_i(t,\vec{p})$ ($i=1,2$) is a spatial Fourier transformation of
$H_i(t,\vec{x})$.
Using the spatial volume average, we can obtain a precise lattice data for both
connected ({\em type1}) and disconnected ({\em type2}) diagrams.

\subsection{The infinite volume reconstruction and QED$_{\mathrm{L}}$ method}

As pointed out earlier, the calculation of $\pi^-\to\pi^+ee$ transition is
similar with the one of QED self energy contributions to $\pi^+$-$\pi^0$
mass difference. So we adopt both methods of QED$_{\mathrm{L}}$ and IVR in our
analysis, although no
QED effects are involved here. Note that a 
particularity of double beta decay is that the $J_{\mu L}$ interpolating operator involves
both vector (V) and axial-vector (A) currents. 
The VA+AV contributions vanish due to the parity symmetry.
After combining the VV and AA contributions, we obtain a large cancellation 
in $\pi^-\to\pi^+ee$ transition amplitude and a significant reduction in its uncertainty.
As a result, the finite-volume (FV) effects are enhanced compared to the
statistical errors.

\subsubsection{QED$_L$}
We start the calculation of the normalized amplitude $A$ defined in
Eq.~(\ref{eq:norm_amplitude}) using the QED$_\mathrm{L}$ method, 
first introduced in Ref.~\cite{Hayakawa:2008an}
\begin{equation}
\begin{split}
    A=A_{\QEDL}+\delta_{\QEDL}(L)=-\frac{1}{F_\pi^2}
    \int_{VT} d^4x\,H_\text{lat}(x)S_\text{lat}(x)
    +\delta_{\QEDL}(L),
\end{split}
\end{equation}
where the integral $\int_{VT}d^4x$ indicates that the integral is performed
within a space-time volume $V\times T$. $H_{\lat}(x)$ is a lattice version of
$H(x)$ function defined in Eq.~(\ref{eq:had_func}) and the scalar propagator is
given by
$S_\text{lat}(x)=\frac{1}{VT}\sum_{p_0}\sum_{\vec{p}\neq\vec{0}}\hat{S}_\text{lat}(p)e^{ipx}$
with $\hat{S}_\text{lat}(p)=\frac{1}{\sum_i\hat{p}_i^2}$ and
$\hat{p}_i=2\sin(p_i/2)$.
Note that the zero mode has been removed from $S_\text{lat}(x)$.
The corresponding FV effects
$\delta_{\QEDL}(L)$ are known to be power-law suppressed. The contribution from $H_0$ is automatically
removed as it is associated with the zero mode of neutrino propagator.

For the VV part of amplitude, $O(1/L)$ and $O(1/L^2)$ corrections in
$\delta_{\QEDL}(L)$ come from the $\pi$ intermediate state. 
These corrections are universal and known as~\cite{Davoudi:2014qua,Borsanyi:2014jba}
\begin{equation}
\label{eq:FV_QEDL}
\begin{split}
    \delta_{\QEDL}^{VV,LO}(L)=\frac{1}{F_\pi^2}\frac{m_\pi c_1}{4\pi L},\quad
    \delta_{\QEDL}^{VV,NLO}(L)=\frac{1}{F_\pi^2}\frac{c_1}{2\pi L^2},
\end{split}
\end{equation}
where $c_1=2.83729$. 
For the AA part, the FV corrections arise from the excited states, e.g.
$\pi\pi$, and
are at the order of $O(1/L^2)$. 
These contributions are not described by the scalar QED and cannot be simply
given by a function of $m_\pi$ and $L$.
So in our analysis we only consider the $O(1/L)$ or partially $O(1/L^2)$ FV corrections 
given in Eq.~(\ref{eq:FV_QEDL}).

\subsubsection{Infinite volume reconstruction}

The detailed description of the IVR method have been given in Ref.~\cite{Feng:2018qpx}, where the
time integral is split into the range of $|t|>t_s$ and $|t|<t_s$
\begin{equation}
\label{eq:IVR}
\begin{split}
A&=A(|t|<t_s)+A(|t|>t_s)\\
    &=-\frac{1}{F_\pi^2}\left(\int_{|t|<t_s} d^4 x\, H'(x)S(x)+\int_{|t|>t_s} d^4 x\, H'(x)S(x)\right)\\
    &=-\frac{1}{F_\pi^2}\left(\int_{|t|<t_s} d^4 x\, H'(x)S(x)+\int d^3 x\,
    H'(t_s,\vec{x})L(t_s,\vec{x})\right)\\
    &=-\frac{1}{F_\pi^2}\left(\int_{V,|t|<ts} d^4 x\, H_\text{lat}'(x)S(x)+\int_V d^3 x\,
    H_\text{lat}'(t_s,\vec{x})L(t_s,\vec{x})\right)+\delta_{\IVR}(L)\\
&\equiv A_{\IVR}+\delta_{\IVR}(L).
\end{split}
\end{equation}
Here a time $t_s$ ($t_s\lesssim L$) is chosen to be sufficiently large for the
intermediate $\pi$-state saturation.
As a result the hadronic function $H'(t,\vec{x})$ at $|t|>t_s$ can be
related to $H'(t_s,\vec{x})$ using the ground-state dominance. Therefore the integral of
$\int_{|t|>t_s} d^4 x\, H'(x)S(x)$ in the second line of Eq.~(\ref{eq:IVR}) can
be written as
$\int d^3 x\, H'(t_s,\vec{x})L(t_s,\vec{x})$ in the third line, with the weighting function $L(t_s,\vec{x})$
given by
\begin{equation}
\begin{split}
L(t_s,\vec{x})&=\frac{1}{(2\pi)^2}\int_0^\infty
    dp\frac{\sin(p|\vec{x}|)}{(E_p+p-m_\pi)|\vec{x}|}e^{-pt_s}.
\end{split}
\end{equation}
In the fourth line of Eq.~(\ref{eq:IVR}) the spatial integrals in the infinite
volume $\int d^3x$ are replaced by the finite-volume integral $\int_V
d^3x$. Besides, the hadronic function $H'(x)$ in the integrand are replaced by
the finite-volume lattice data $H_{\lat}'(x)$.
According to the above changes, the FV corrections $\delta_{\IVR}(L)$ 
account for two effects: 1) the difference between
$H_{\lat}'(x)$ and $H'(x)$ inside the spacetime box
where lattice data are available, 2) the integral outside the spatial volume
$V$, where lattice data are not available. Both effects
are exponentially suppressed as demonstrated in Ref.~\cite{Feng:2018qpx}. 
Thus for sufficiently large volume, we can ignore $\delta_{\IVR}(L)$. 

\begin{figure}[htbp]
	\centering
	\includegraphics[width=0.8\textwidth]{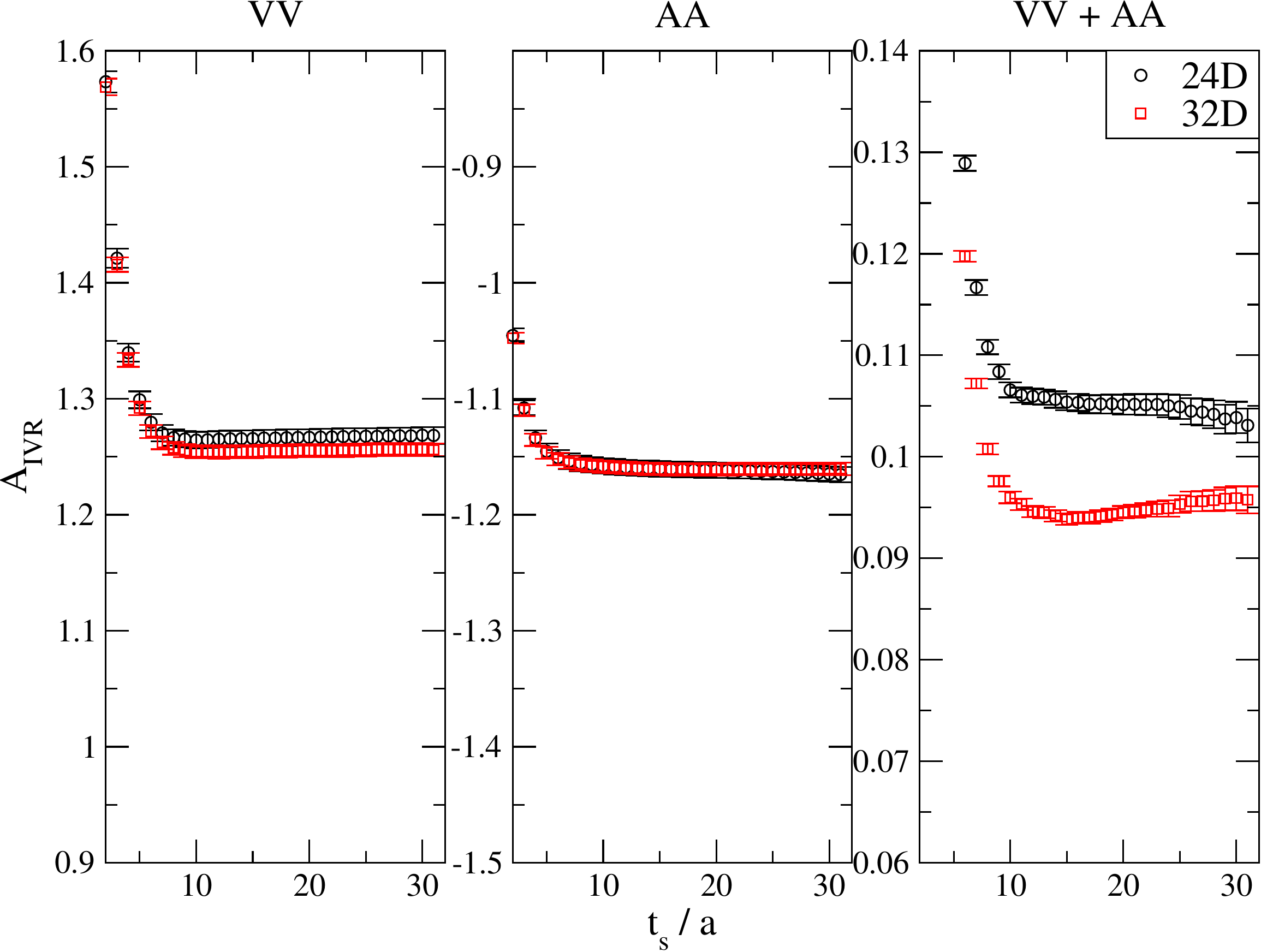}
    \caption{The amplitude $A_{\IVR}$ as a function of $t_s$ for ensembles 24D (black
    circle) and 32D (red cubic).
    The left, middle and right panels show the results for the VV, AA and VV+AA
    contributions, respectively. 
    Due to the cancellation between VV and AA
    contributions, the finite-volume
    effects in the combined results become significant when compared to the
    statistical errors.} 
    \label{fig:enhanced_FV}
\end{figure}

In this calculation, the lattice volumes listed in Table~\ref{tab:ensemble_parameter} are relatively
small. 
Besides, FV effects are enhanced due to the cancellation
between VV and AA contributions. As a
consequence, even exponentially suppressed, the size of $\delta_{\IVR}(L)$ are statistically
significant. This can be confirmed by Fig.~\ref{fig:enhanced_FV}, where we compare the
amplitude $A_{\IVR}$ for ensembles 24D and 32D. 
Although VV and AA parts of the amplitudes are nearly consistent
for the two ensembles, a significant discrepancy is found 
after the VV and AA parts are combined together.

\begin{figure}[htbp]
	\centering
	\includegraphics[width=0.8\textwidth]{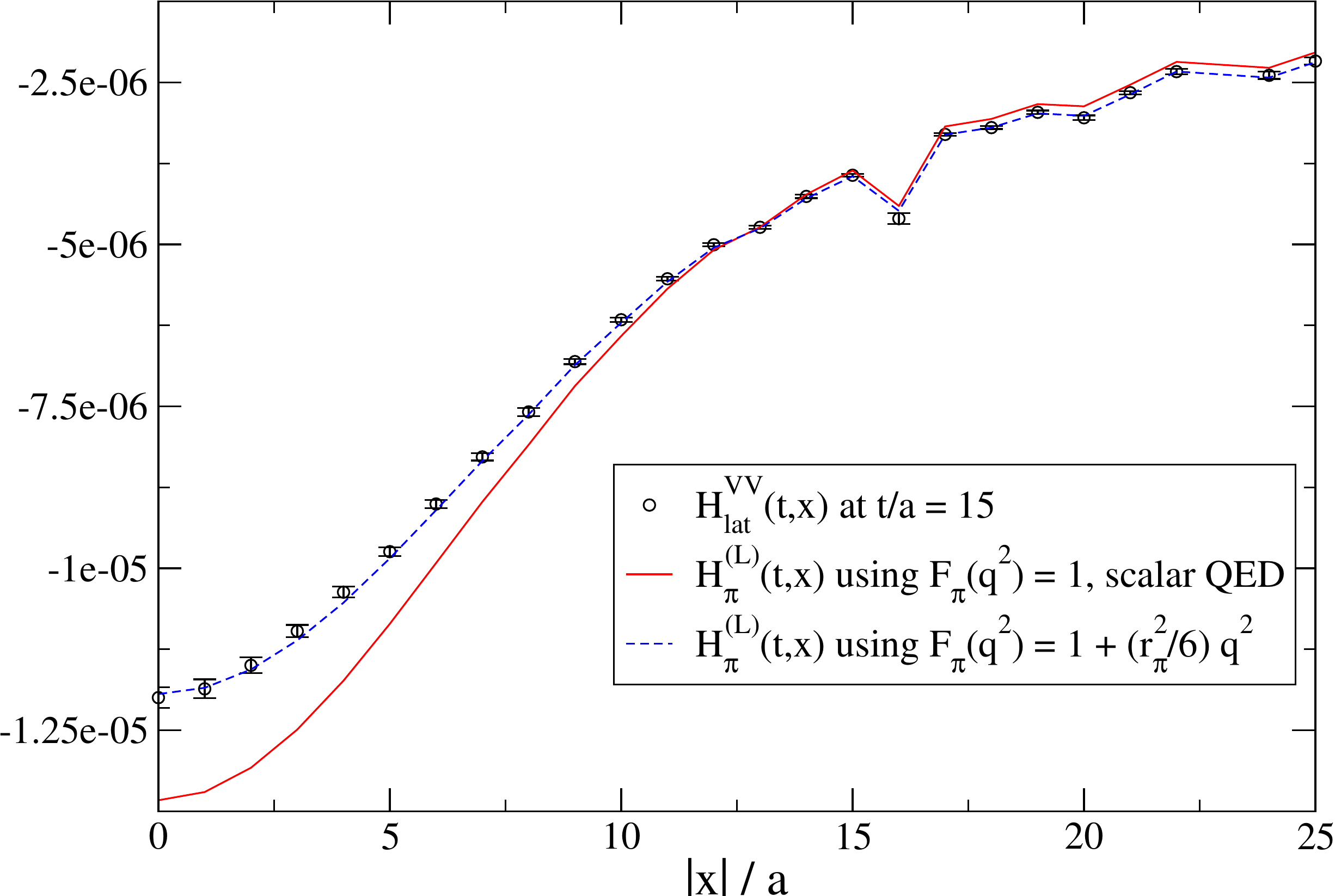}
    \caption{The VV part of the hadronic function $H_\text{lat}^{VV}(t,\vec{x})$ at $t/a=15$
    as a function of $|\vec{x}|$. A comparison is made between lattice data of $H_\text{lat}^{VV}(t,\vec{x})$ and 
    the pion contribution $H_\pi^{(L)}(t,\vec{x})$ by setting $F_\pi(q^2)=1$ and
    $F_\pi(q^2)=1+(r_\pi^2/6)q^2$.}
    \label{fig:LD_contribution}
\end{figure}

Due to the enhanced FV effects,
it is necessary to estimate the size of
$\delta_{\IVR}(L)$ in our calculation.
Note that $\delta_{\IVR}(L)$ receives the dominant contributions from pion intermediate
states. The relevant hadronic function can be written as
\begin{equation}
\begin{split}
H_\pi(x)&=\int\frac{d^3 p}{(2\pi)^3 2E_{\pi,\vec{p}}} \langle
    \pi^+(0)|J_{\mu L}(0)|\pi^0(p)\rangle\langle \pi^0(p)|J_{\mu
    L}(0)|\pi^-(0)\rangle e^{i\vec{p}\cdot\vec{x}}e^{-(E_{\pi,\vec{p}}-m_\pi)|t|}\\
&=-\int\frac{d^3 p}{(2\pi)^3 2E_{\pi,\vec{p}}}
    m_\pi(m_\pi+E_{\pi,\vec{p}})\left[F_\pi(q^2)\right]^2
    e^{i\vec{p}\cdot\vec{x}}e^{-(E_{\pi,\vec{p}}-m_\pi)|t|},
\end{split}
\end{equation}
where $F_\pi(q^2)$ is the pion form factor with
$q^2=2m_\pi(m_\pi-E_{\pi,\vec{p}})$ and $E_{\pi,\vec{p}}=\sqrt{m_\pi^2+\vec{p}^2}$.
The corresponding hadronic function in the finite volume is then given by
\be
H_\pi^{(L)}(x)
=-\frac{1}{L^3}\sum_{\vec{p}} \frac{1}{2E_{\pi,\vec{p}}}
    m_\pi(m_\pi+E_{\pi,\vec{p}})\left[F_\pi(q^2)\right]^2
    e^{i\vec{p}\cdot\vec{x}}e^{-(E_{\pi,\vec{p}}-m_\pi)|t|}.
\ee
Using the pion contribution as input, we can approximate
$\delta_{\IVR}(L)$ by
\begin{equation}
\label{eq:pi_contribution}
    \delta_{\IVR}(L)\approx\delta_{\IVR}^\pi(L)= A^\pi_{\IVR}-A^{\pi}_{\IVR}(L)\\
\end{equation}
where $A^\pi_{\IVR}$ and $A^{\pi}_{\IVR}(L)$ are determined using the hadronic
functions
$H_\pi(x)$ and $H_\pi^{(L)}(x)$, respectively.
Depending on the functional forms of $F_\pi(q^2)$, we have two estimates for
$\delta_{\IVR}^{\pi}$.
\begin{itemize}
\item In the framework of scalar QED, we set $F_\pi(q^2)=1$, where the internal
electromagnetic
structure of pion is neglected. As shown in Fig.~\ref{fig:LD_contribution}, at
large $t$ and $\vec{x}$, e.g. $t/a=15$ and $|\vec{x}|/a>12$, the lattice results of the
VV hadronic function $H_\text{lat}^{VV}(x)$ agree relatively well with the $H_\pi^{(L)}(x)$
from scalar QED. We denote the $\delta_{\IVR}^{\pi}$ using $F_\pi(q^2)=1$ as
        $\delta_{\IVR}^{\pi,(1)}$.
\item If we adopt the expression of $F_\pi(q^2)=1+(r_\pi^2/6)q^2$ and use the PDG
    value of the pion charge radius $r_\pi=0.659(4)$ fm as input~\cite{Tanabashi:2018oca}, 
a better consistency is found between the lattice data of $H_\text{lat}^{VV}(x)$
and $H_\pi^{(L)}(x)$ at $t/a=15$. In this case, $\delta_{\IVR}^{\pi}$ is denoted
as $\delta_{\IVR}^{\pi,(2)}$.
\end{itemize}
From Fig.~\ref{fig:LD_contribution} we confirm that the long distance behavior of $H_\text{lat}^{VV}(x)$
can be well described by the $\pi$ intermediate state. Since FV effects mainly come from long distance 
physics, $\delta_{\IVR}^\pi(L)$ can provide a good estimate for $\delta_{\IVR}(L)$.
It is worthwhile to point out that both $\delta_{\IVR}(L)$ and $\delta_{\IVR}^\pi(L)$ are exponentially
suppressed as $L$ increases, thus the residual FV effects in
$\delta_{\IVR}(L)-\delta_{\IVR}^\pi(L)$ are also exponentially suppressed and
thus well under control.

\section{Numerical results}

\begin{figure}[htbp]
	\centering
	\includegraphics[width=0.8\textwidth]{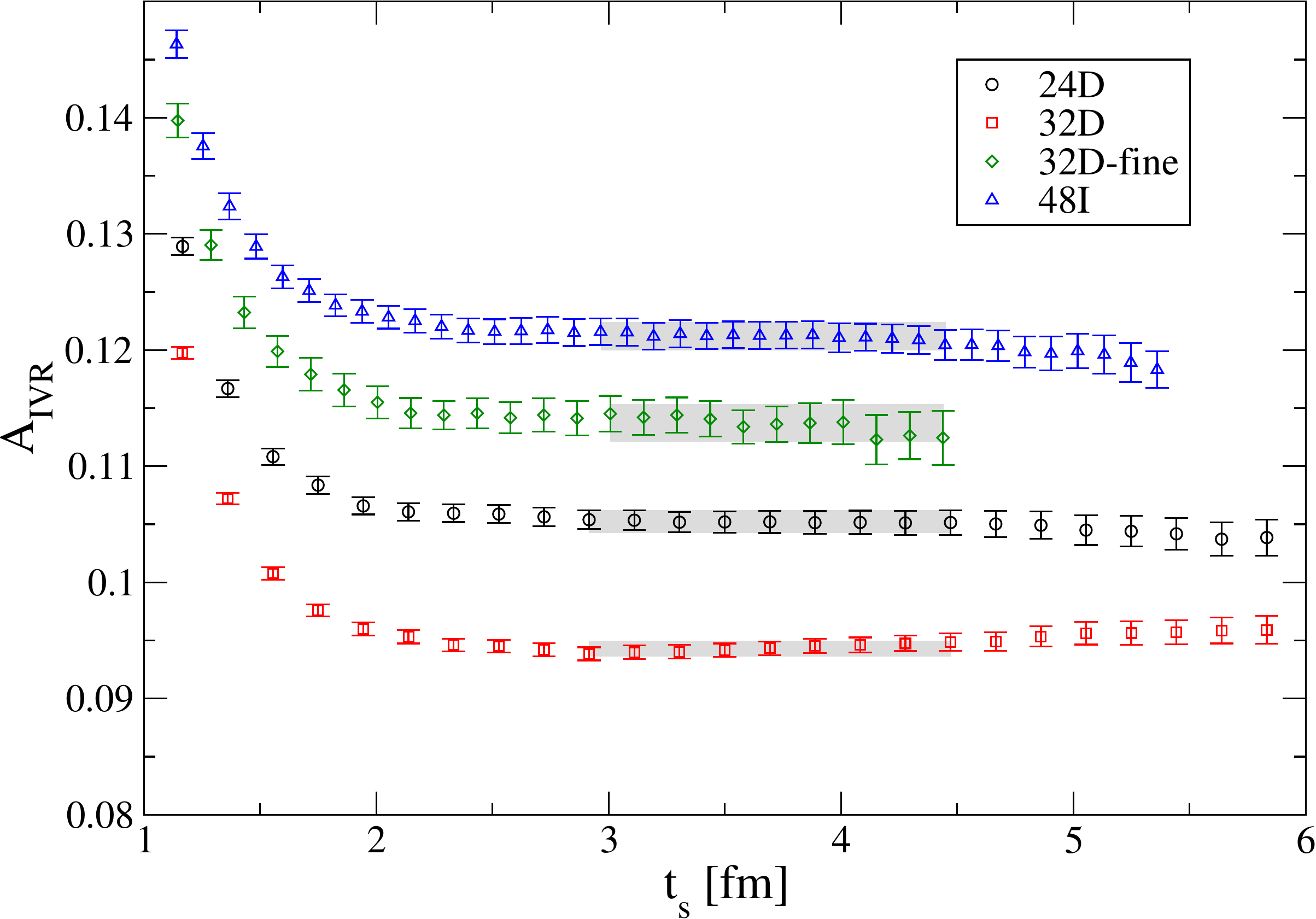}
    \caption{The amplitude $A_{\IVR}$ as a function of $t_s$.}
    \label{fig:IVR}
\end{figure}

The IVR amplitudes $A_{\IVR}$ as a function of $t_s$ are shown in
Fig.~\ref{fig:IVR} together with a fit to a constant. 
All the ensembles shown in Fig.~\ref{fig:IVR} visibly agree with the corresponding 
fit in the window of $3$ fm $\lesssim t_s\lesssim$ $4.5$ fm 
and lead to reasonable values of $\chi^2$ per degree of freedom.
The fitting results are shown in
Table~\ref{tab:IVR}.

\begin{table}[htbp]
	\small
	\centering
	\begin{tabular}{c|ccc|ccc}
		\hline\hline
        Ensemble& $A_{\IVR}$ & $A_{\IVR}+\delta_{\IVR}^{\pi,(1)}$ &
        $A_{\IVR}+\delta_{\IVR}^{\pi,(2)}$ & $g_{\nu}^{\pi\pi,(0)}$ &
        $g_{\nu}^{\pi\pi(1)}$ & $g_{\nu}^{\pi\pi,(2)}$\\
		\hline
        24D & 0.1052(9) & 0.0872(9) & 0.0841(10) & $-10.63(6)$ & $-12.14(6)$ &
        $-12.46(7)$\\
        32D & 0.0943(6) &0.0864(6) & 0.0854(6) & $-11.53(4)$ & $-12.19(4)$ &
        $-12.28(5)$\\
        32D-fine &0.1137(15)& 0.0951(15) & 0.0913(14) & $-10.04(12)$ &
        $-11.57(12)$ & $-11.88(12)$ \\
        48I &0.1212(12)& 0.1100(11) & 0.1071(12) & $-9.27(7)$ & $-10.22(7)$ &
        $-10.47(8)$\\
		\hline
	\end{tabular}%
    \caption{Results of amplitude $A_{\IVR}$ and low energy constant
    $g_\nu^{\pi\pi}(\mu)$ at $\mu=m_\rho$ for four ensembles. The three columns $A_{\IVR}$,
    $A_{\IVR}+\delta_{\IVR}^{\pi,(1)}$ and $A_{\IVR}+\delta_{\IVR}^{\pi,(2)}$
    correspond to
    the amplitude without FV correction and the ones with corrections
    $\delta_{\IVR}^{\pi,(1)}$ and $\delta_{\IVR}^{\pi,(2)}$. The
    values of $g_\nu^{\pi\pi,(0)}$, $g_\nu^{\pi\pi,(1)}$ and $g_\nu^{\pi\pi,(2)}$
are obtained by putting $A_{\IVR}$, $A_{\IVR}+\delta_{\IVR}^{\pi,(1)}$ and
    $A_{\IVR}+\delta_{\IVR}^{\pi,(2)}$ into Eq.~(\ref{eq:A_ChPT}), respectively.} 
	\label{tab:IVR}%
\end{table}%

\subsection{Finite-volume effects}

For ensembles 24D ($m_\pi L=3.3$) and 32D ($m_\pi
L=4.5$), the results $A_{\IVR}$ disagree by $\sim10\%$, which is much larger than
their statistical errors. We evaluate the FV 
corrections $\delta_{\IVR}^{\pi,(1)}$ and
$\delta_{\IVR}^{\pi,(2)}$ at $t_s\simeq 3.75$ fm by adopting Eq.~(\ref{eq:pi_contribution}) 
and using the input of
$F_\pi(q^2)=1$ and $F_\pi(q^2)=1+(r_\pi^2/6)q^2$, respectively.
As can be seen in Table~\ref{tab:IVR}, after adding the corrections, 
the large discrepancy between 24D and 32D
results vanishes. For the ensembles with smallest volume, e.g. 24D and
32D-fine, the results for $A_{\IVR}+\delta_{\IVR}^{\pi,(1)}$ and
$A_{\IVR}+\delta_{\IVR}^{\pi,(2)}$ still differ by 4\%, suggesting that the FV
effects at the level of $\delta_{\IVR}^{\pi,(2)}-\delta_{\IVR}^{\pi,(1)}$ 
shall be taken into account in the error budget.

\begin{figure}[htbp]
	\centering
	\includegraphics[width=0.8\textwidth]{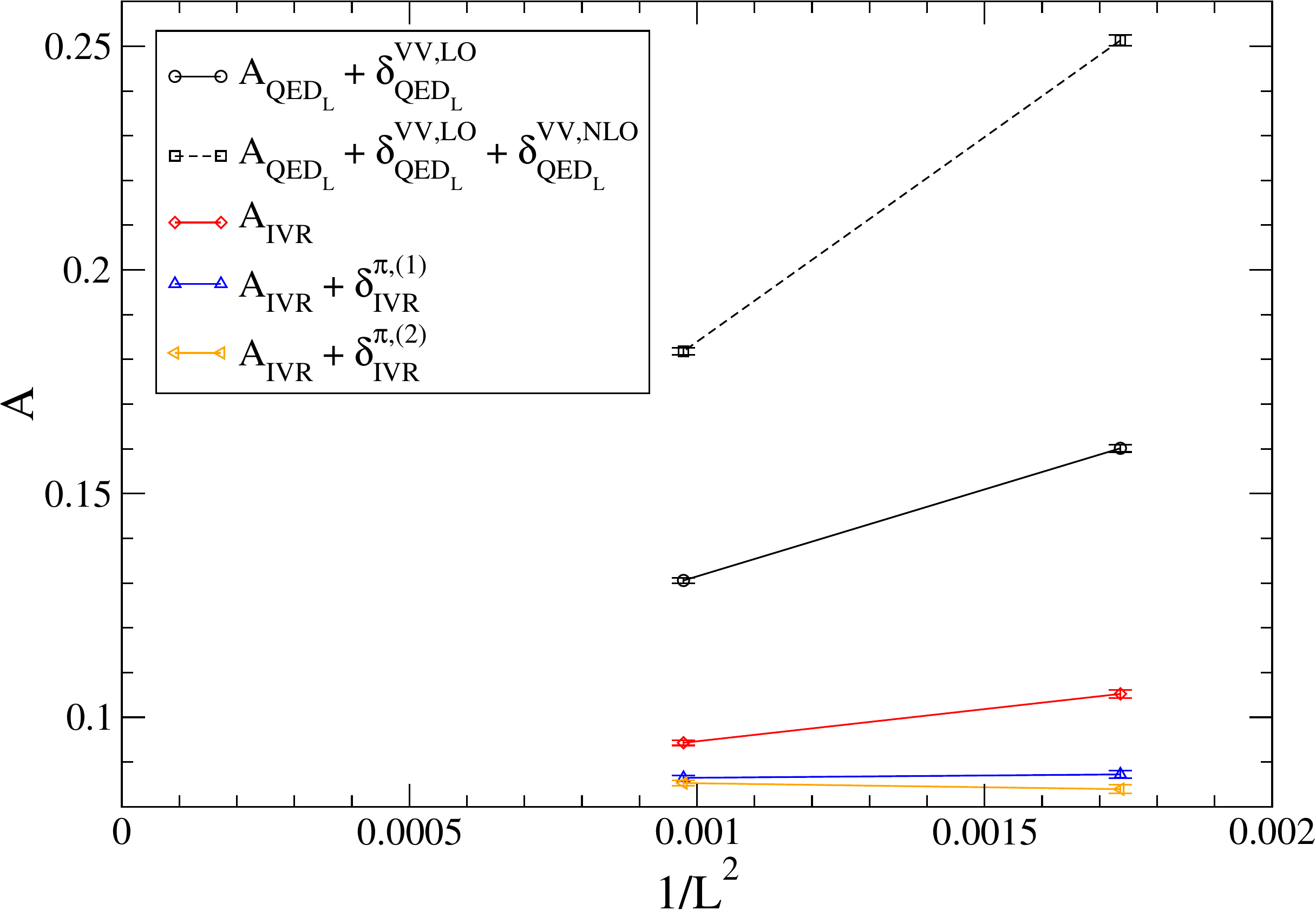}
    \caption{QED$_{\mathrm{L}}$ and IVR results of amplitude $A$ for ensembles 24D and 32D. The black
    circle and square data are obtained from QED$_\mathrm{L}$ method with LO and
    partially NLO FV corrections given in Eq.~(\ref{eq:FV_QEDL}). The red diamond, blue
    triangle-up and orange triangle-left data are obtained using IVR method,
    indicating the amplitude $A_{\IVR}$, $A_{\IVR}+\delta_{\IVR}^{\pi,(1)}$ and
    $A_{\IVR}+\delta_{\IVR}^{\pi,(2)}$, respectively.}
    \label{fig:FV_corr}
\end{figure}

In Fig.~\ref{fig:FV_corr} we compare the results for ensemble 24D and 32D from
QED$_\mathrm{L}$ and IVR methods.
With LO and partially NLO FV corrections, the results of $A_{\QEDL}+\delta_{\QEDL}^{VV,LO}$
and $A_{\QEDL}+\delta_{\QEDL}^{VV,LO}+\delta_{\QEDL}^{VV,NLO}$ still have
significant dependence on lattice volumes; while for the IVR results, such
dependence is very mild. We therefore use the IVR results in
the final analysis. 

\subsection{Continuum extrapolation}

\begin{figure}[htbp]
	\centering
	\includegraphics[width=0.8\textwidth]{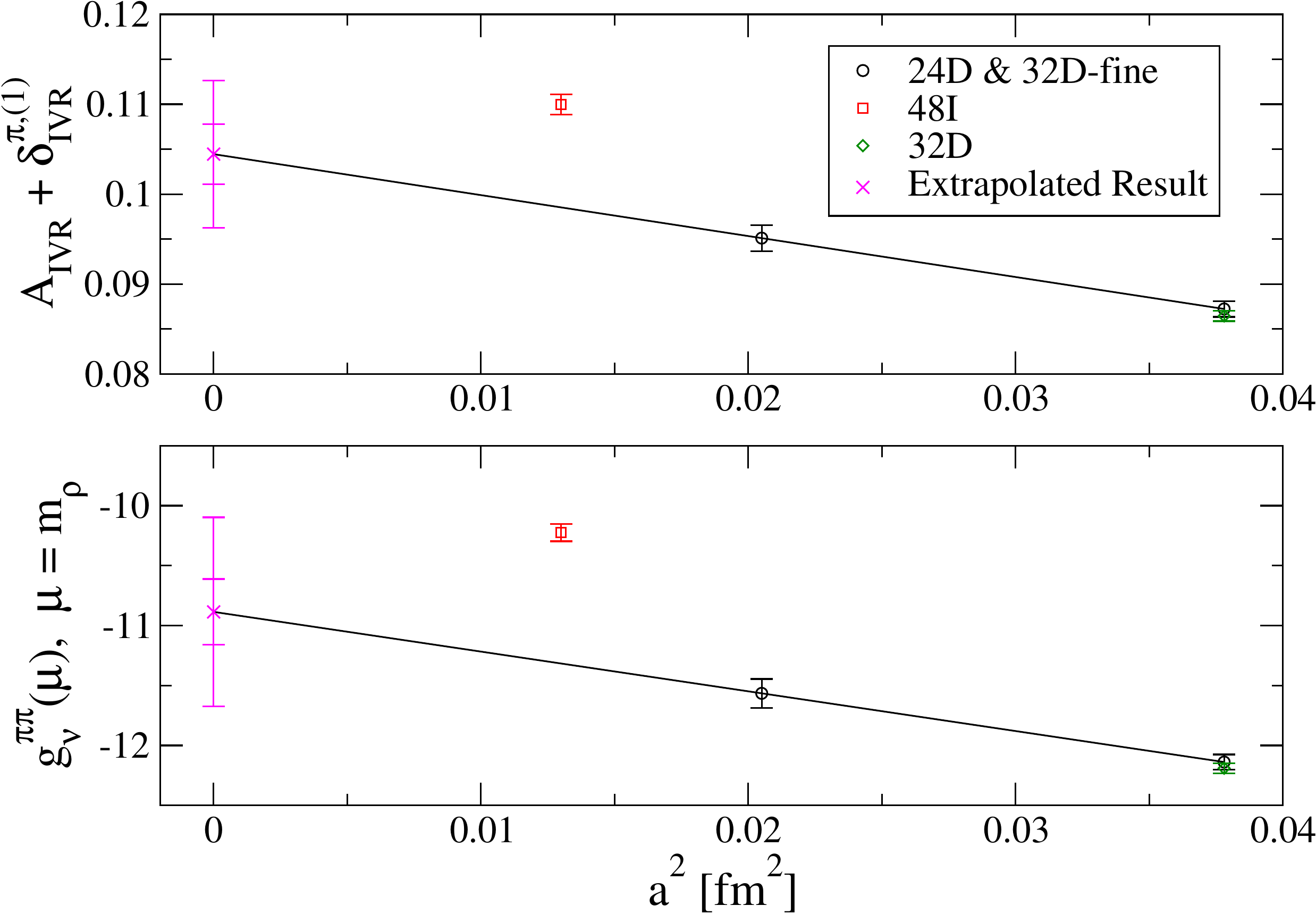}
    \caption{The amplitude $A_{\IVR}+\delta_{\IVR}^{\pi,(1)}$ and the
    corresponding
    low energy constant $g_\mu^{\pi\pi}(\mu)$ at $\mu=m_\rho$ as a
    function of lattice spacing square.}
    \label{fig:cont_extrap}
\end{figure}

For all the four ensembles,
the amplitude $A_{\IVR}+\delta_{\IVR}^{\pi,(1)}$ as a function of lattice spacing square is shown
in Fig.~\ref{fig:cont_extrap}. 
We use the results from 24D ($a^{-1}=1.015$ GeV) and 32D-fine ($a^{-1}=1.378$ GeV) ensembles to perform a continuum
extrapolation and obtain $A_{\IVR}^{\text{cont}}=0.1045(34)$ at the continuum limit. 
We have another ensemble 48I with $a^{-1}=1.73$ GeV
to
further examine the lattice artifacts. As
the 48I ensemble is simulated with Iwasaki gauge action, it contains the
different lattice artifacts compared to the 24D and 32D-fine ensembles, where
Iwasaki+DSDR action is used. Although the 48I result cannot be used in the
continuum extrapolation directly, it helps us to estimate the size of
the lattice artifacts.

\subsection{Results}

As we only use two ensembles for continuum extrapolation, the residual lattice
artifacts might not be fully controlled by the extrapolation.
To be conservative we
quote the difference between the extrapolated result $A_{\IVR}^\text{cont}$ and the 48I
result $A_{\IVR}^\text{48I}$ as the size
of systematic effects, namely
$\delta_a=|A_{\IVR}^\text{cont}-A_{\IVR}^\text{48I}|=0.0055$.
In Fig.~\ref{fig:cont_extrap} the amplitude at the continuum limit is
obtained using
$A_{\IVR}+\delta_{\IVR}^{\pi,(1)}$.
We also
calculate $A_{\IVR}+\delta_{\IVR}^{\pi,(2)}$ using $F_\pi(q^2)=1+(r_\pi^2/6)q^2$.
The $O(q^2)$ term in $F_\pi(q^2)$ causes a shift $\delta_L=0.0050$ in the
amplitude $A_{\IVR}^\text{cont}$. Such effect is included as a systematic uncertainty for the residual
FV effects. To sum up, the final result for the amplitude defined in
Eq.~(\ref{eq:norm_amplitude}) is given by
\be
\label{eq:A_results}
A=0.1045(34)(50)_{L}(55)_{a},
\ee
where the first uncertainty is statistical, the second and third ones are the
systematic errors for finite volume and lattice artifacts. 
Putting the amplitude into Eq.~(\ref{eq:A_ChPT}), we obtain the results for low energy constant
$g_\nu^{\pi\pi}(\mu)$ at $\mu=m_\rho$ with $m_\rho$ the rho meson mass. The
values of $g_\nu^{\pi\pi,(0)}$, $g_\nu^{\pi\pi,(1)}$ and $g_\nu^{\pi\pi,(2)}$
are obtained using $A_{\IVR}$, $A_{\IVR}+\delta_{\IVR}^{\pi,(1)}$ and
$A_{\IVR}+\delta_{\IVR}^{\pi,(2)}$ as inputs, respectively. These results are put
in Table~\ref{tab:IVR}.
Following the similar procedure described above, the final result for $g_\nu^{\pi\pi}$ with both
statistical and systematic uncertainties is given by
\be
\label{eq:g_results}
g_\nu^{\pi\pi}(\mu)\Big|_{\mu=m_\rho}=-10.89(28)(33)_L(66)_a.
\ee

\section{Conclusion}

We perform a lattice QCD calculation of the
amplitude of the neutrinoless double beta decay $\pi^-\to\pi^+ ee$.
The hadronic function $H_0(x)$, which contains the contributions 
from vacuum state, are subtracted. Such
subtraction
removes the exponentially growing terms in the Euclidean time integral. The
remaining hadronic function $H'(x)=H(x)-H_0(x)$ can be used to determine the
normalized amplitude $A=\mathcal{A}^M/\mathcal{A}_0^M-1$, which can be considered as
a fractional deviation between the total decay amplitude $\mathcal{A}^M$ and the
leading-order $\chi$PT predication $\mathcal{A}_0^M$.

In the calculation, we find large FV effects in the decay amplitude.
By comparing two approaches, QED$_{\mathrm{L}}$ and IVR, 
we finally adopt the IVR method in our study, as the associated FV effects are
exponentially suppressed and much smaller than that from QED$_{\mathrm{L}}$.
By adding the FV corrections $\delta_{\IVR}^{\pi,(1)}$ or $\delta_{\IVR}^{\pi,(2)}$
contributed by the pion intermediate state, the 24D and 32D results become
consistent. The residual FV effects are estimated by the size of
$\delta_{\IVR}^{\pi,(1)}-\delta_{\IVR}^{\pi,(2)}$. 
After the continuum extrapolation, the final result of $A$ is given in Eq.~(\ref{eq:A_results}),
with a $\sim3\%$ statistical error. For the systematic effects including both FV
effects and lattice artifacts, we estimate them at the level of $\sim$10\%.

By putting the amplitude $A$ into the $\chi$PT formula, we
determine the low energy constant $g_\nu^{\pi\pi}$ in Eq.~(\ref{eq:g_results}). This result
is close to the $g_\nu^{\pi\pi}$ from $\pi^-\pi^-\to ee$ decay, which is given
in Eq.~(\ref{eq:LEC1}), suggesting that $\chi$PT works well in the pion
sector. It has been found in Ref.~\cite{Cirigliano:2018hja} that 
a leading-order,
short-range contribution needs to be introduced in the $nn\to pp ee$ decay, which 
breaks down Weinberg's power-counting scheme. Moving the calculation from the
pion to the nucleon sector is the
next step of our $0\nu2\beta$ project.
It is interesting to examine the impact of this short-range contribution in our future study.

\begin{acknowledgments}

We gratefully acknowledge many helpful discussions with our colleagues from the
RBC-UKQCD Collaboration. We warmly thank N.~H.~Christ, V. Cirigliano, W.~Dekens, W. Detmold,
E.~Mereghetti, D.~Murphy, A. Nicholson, U.~van Kolck and A. Walker-loud for useful discussion.
X.F. and X.-Y.T. were supported in part by NSFC of China under Grant No. 11775002.
L.C.J acknowledge support by DOE grant DE-SC0010339.
The computation was performed under the ALCC Program of
the US DOE on the Blue Gene/Q (BG/Q) Mira computer at the Argonne Leadership Class Facility,
a DOE Office of Science Facility supported under Contract DE-AC02-06CH11357.
Part of the computation was carried out on facilities of the USQCD
Collaboration, which are funded by the Office of Science of the U.S. Department
of Energy.
The calculation was also carried out on TianHe-1 (A) at Chinese National Supercomputer Center in Tianjin.
\end{acknowledgments}

\bibliography{paper}

\end{document}